\documentclass[twocolumn,showpacs,amsmath,amssymb,prb,graphics,graphicx]
{revtex4}
  \usepackage{graphicx}
\usepackage{bm}
\usepackage{dcolumn}
\begin{document}
\title{Penetration depth anisotropy in two-band superconductors}
\author{V. G. Kogan, N. V. Zhelezina}
  \affiliation{
     Ames Laboratory - DOE and Department of Physics,
   Iowa State University, Ames, IA 50011}
\date{July, 2003; \today}
\begin{abstract}
The anisotropy of the London penetration depth is evaluated for two-band
superconductors with arbitrary inter- and intra-band scattering times. If one
of the bands is clean and the other is dirty in the absence of
inter-band scattering, the anisotropy is dominated by the Fermi surface of the
clean band and is weakly temperature dependent. The inter-band scattering also
suppress the temperature dependence of the anisotropy.
    \end{abstract}
\maketitle

%{\it Introduction.}

The two-gap superconductivity of MgB$_2$  is  established
experimentally \cite{Bouquet,Szabo,Giubileo,Junod,Schmidt} and
by solving the Eliashberg equations for the gap distribution
on the Fermi surface.\cite{Choi,Mazin} According to the latter, the gap on the
four Fermi surface sheets of this material has two sharp maxima:
$\Delta_1\approx 1.7\,$meV at the two   $\pi$-bands  and
$\Delta_2\approx 7\,$meV at the two $\sigma$-bands. Within each of
these groups, the spread of the gap values is small.

A number of physical properties of MgB$_2$ were reasonably well
described within a
model with two constant gaps on two separate Fermi sheets.
Still, the data on  anisotropy of the magnetic field penetration depth
$\lambda$ are  controversial. The anisotropy parameter
$\gamma_\lambda=\lambda_c/\lambda_a$ has been calculated within the
{\it weak-coupling clean-limit} model and   shown to increase from about 1.1 at
$T=0$ to $\approx 2.6$ at   $T_c$. \cite{Kogan} Similar prediction has been
made within Eliashberg formalism. \cite{Golubov} Qualitatively, the
predictions
were confirmed in STM, \cite{STM,Morten} small angle neutron
scattering, \cite{Cubitt}  and magnetization experiments. \cite{Lyard}
However, other groups recorded different behavior.
\cite{Jap-torque,Angst,Caplin}   Given variety of samples used, it seems
imperative to consider   effects of scattering upon $\gamma_\lambda$,
a non-trivial
problem given different roles of the intra- and inter-band scattering in
   two-band  materials.

In the following, we reiterate that in the presence of inter-band scattering,
the energy gap in the quasiparticle excitation spectrum as revealed
by the density
of states (DOS) differs from both $\Delta_1 $ and $\Delta_2$,
\cite{Mosk1,Schop}
the situation reminiscent of the Abrikosov-Gor'kov pair breaking. \cite{AG}
For this reason, we use the term ``order parameter" for $\Delta$'s.
We stress that
    all thermodynamic properties depend on the actual DOS and  are affected
by the inter-band scattering. Then, we show that
$\gamma_\lambda=\lambda_c/\lambda_{ab}$  depends  on both inter- and intra-band
scattering.

   Our approach is based on the quasiclassical version of the BCS theory for
a general anisotropic Fermi surface and for an arbitrary anisotropic order
parameter  $\Delta ({\bm k})$. \cite{Eil} In the absence of currents
and fields we
have for the Eilenberger Green's functions $f({\bm k},\omega)$ and $g({\bm
k},\omega)$:
\begin{equation}
0=2\Delta g   -2\omega f+I \,,\qquad
\label{eil1}
1=g^2+ f^2 \,.
\end{equation}
Here the scattering term $I$ is given by the integral over the full Fermi
surface:
\begin{equation}
I( {\bm k} )=\int d^2{\bm q}\,\rho({\bm q})\,W({\bm k},{\bm
q})\,[g({\bm k})f({\bm
q})-f({\bm k})g({\bm q})]  \label{I}
\end{equation}
with $W({\bm k},{\bm q})$ being the scattering probability from   $ {\bm q}$
to  $ {\bm k}$ at the Fermi surface.  The Matsubara frequencies are $\omega=\pi
T(2n+1)$ with an integer $n$ ($\hbar=1$). The {\it local} DOS
$\rho({\bm q})$ is
normalized: $\int d^2{\bm q}\,\rho({\bm q})=1$.

This system (\ref{eil1})-(\ref{I}) should be complemented with an
equation  for
$\Delta( {\bm k})$. We will not use it here, rather taking $\Delta(
{\bm k})$ as a
given.  This simplifies the problem greatly because solving for   $\Delta( {\bm
k})$ usually involves a number of assumptions which are difficult to control.

  We use  the approximation of the scattering time $\tau$:
\begin{equation}
\int d^2{\bm q}\,\rho({\bm q})\,W({\bm k},{\bm q})\, f({\bm
q})=\langle f\rangle/\tau\,; \label{tau}
\end{equation}
$\langle ...\rangle$  stands for the average over the  Fermi surface.
Clearly, the approximation amounts to the scattering probability
$W=1/\tau$ being constant for any  ${\bm k}$ and ${\bm q}$.

For two well-separated Fermi surface sheets, the probabilities of intra-band
scatterings may differ from each other and from processes involving
${\bm k}$ and
${\bm q}$ from different bands. The effects of the inter-
  and intra-band scattering upon various properties of the system are
different, e.g., the intra-band scattering does not affect $T_c$, whereas the
inter-band does.  Therefore, Eq. (\ref{tau}) is replaced by
  \begin{equation}
\int d^2{\bm q_\alpha}\,\rho({\bm q_\alpha})\,W({\bm k_\beta},{\bm q_\alpha})\,
f({\bm q_\alpha})=\langle f\rangle_\alpha/\tau_{\beta\alpha}\,.
\end{equation}
Here  $\alpha,\beta =1,2$ are  band indices and $\langle ...\rangle_{\alpha}$
denotes averaging only over the $\alpha$-band.

We now assume  the order parameters $\Delta({\bm k}_{\alpha} )$ taking constant
values $\Delta_1$ and $\Delta_2$ at each of the two bands.
Writing Eq.\ (\ref{eil1}) for ${\bm k}$ in the first band, we have:
\begin{eqnarray}
&0&=2\Delta_1 g_1 - 2\omega f_1\nonumber\\ &+& \frac{1}{\tau _{11}}[g_1 \langle
f \rangle_1 -f_1\langle g
\rangle_1]+ \frac{1}{\tau _{12}} [g_1 \langle f
\rangle_2 -f_1\langle g \rangle_2] \,.
\label{eq5}
\end{eqnarray}
For a uniform sample in zero field and with ${\bm k}$ independent $\Delta$'s in
each band, the functions $f,g$ are ${\bm k}$ independent:  $\langle f
\rangle_\alpha = f_\alpha$ and $\langle g \rangle_\alpha=g_\alpha$. Then, we
have:
  \begin{eqnarray}
0=\Delta_1 g_1 - \omega f_1 + [g_1  f_2 -f_1  g_2]/2\tau _{12}\,.
\label{eq7}
\end{eqnarray}
The equation for the second band differs from this by replacement
$1\leftrightarrow
2$. The fact that $\tau _{11}$ and $\tau_{22}$ do not enter the system
(\ref{eq7}) is similar to the case of one-band isotropic material for which
non-magnetic scattering has no effect upon $T_c$ (the Anderson theorem). It
is the inter-band scattering that makes the difference in the two-gap case, the
fact stressed already in early work. \cite{Mosk1,Schop}

  Two equations (\ref{eq7}) are complemented with normalizations
$g_\alpha^2+f_\alpha^2=1$ to form a sufficient set. Following Ref.
\onlinecite{AG}, we introduce variables
$u_\alpha=g_\alpha/f_\alpha$ and obtain after simple
algebra: \cite{Sung,Mosk1,Schop}
\begin{eqnarray}
\frac{\omega}{\Delta_1} = u_1 + \zeta_1 \frac{u_1-u_2}{\sqrt{u_2^2+1}}\,,\qquad
\zeta_1=\frac{1}{2\tau_{12}\Delta_1}
  \,;\nonumber\\
\frac{\omega}{\Delta_2} = u_2 + \zeta_2 \frac{u_2-u_1}{\sqrt{u_1^2+1}}\,,\qquad
\zeta_2=\frac{1}{2\tau_{21}\Delta_2} \,.
\label{b2}
\end{eqnarray}
The Eilenberger functions in terms of  variables $u$ read:
\begin{equation}
f=1/\sqrt{1+u^2}\,,\qquad g=u/\sqrt{1+u^2}\,.
\label{fg}
\end{equation}

%\subsubsection{GL domain}

In general, the system (\ref{b2}) can be solved only numerically. However, near
$T_c$, $u=g/f\gg 1$ and  one obtains:
\begin{eqnarray}
u_1 =\frac{
\omega}{\Delta_1}\,\frac{
\omega+\zeta_1\Delta_1+\zeta_2\Delta_2}{\omega+(\zeta_1+\zeta_2)\Delta_2} \,;
\label{GL}
\end{eqnarray}
$u_2$ is obtained by   $1\leftrightarrow 2$.
Clearly, $u_{\alpha} =\omega/\Delta_{\alpha}$ in the
absence of inter-band scattering.
For   $\zeta\gg 1$, we have:
\begin{equation}
u_1\approx u_2\approx \frac{\omega}{\epsilon^*}\,,\quad
\epsilon^*=\frac{ (\zeta_1+\zeta_2)\Delta_1\Delta_2}{
  \zeta_1\Delta_1+\zeta_2\Delta_2}\,.
\label{GL_limit}
\end{equation}

Moreover, if the inter-band scattering is strong, Eq.\
(\ref{GL_limit}) holds at
any $T$. To see  this,  look for solutions of Eqs.\ (\ref{b2}) in the form
\begin{equation}
  u_\alpha=\frac{\omega}{\epsilon^*}+v_\alpha\,,\qquad \alpha=1,2\,\,,
\label{strong inter}
\end{equation}
where $v_\alpha$ are small corrections. Substitute these in Eqs.\
(\ref{b2}) and keep  only linear terms in $v$ to obtain
\begin{eqnarray}
v_1&=&  \frac{g^* (\epsilon^*-\Delta_1)}{\Delta_1
[1+g^*(\zeta_1+\zeta_2)]}\nonumber\\
&+&\frac{\epsilon^*(\zeta_1\Delta_1+
\zeta_2\Delta_2)-\Delta_1\Delta_2(\zeta_1+\zeta_2)
  }{\Delta_1\Delta_2[1+g^*(\zeta_1+\zeta_2)]}\,,
  \label{v1,strong}
\end{eqnarray}
where $g^*=\omega/\sqrt{\omega^2+\epsilon^{*2}}$.
For $\zeta_\alpha\to\infty$, $v_1$ remains small only if $\epsilon^*$
is given by
   expression (\ref{GL_limit}).

%\subsection{DOS}

The DOS for two-band materials is $N(\epsilon)=N(0)\,{\rm Re}(\nu_1\,g_1 +
\nu_2  \, g_2)_{\omega\to i\epsilon}$;  $\nu_\alpha$ are fractions of the total
DOS $N(0)$ at the two pieces of the Fermi surface. For strong inter-band
scattering, this gives in the lowest approximation
  \begin{equation}
  N(\epsilon)=N(0)\,\frac{\epsilon^*}{\sqrt{\epsilon^2-\epsilon^{*2}}};
  \label{N1+N2}
  \end{equation}
i.e., $\epsilon^*$ is the common for   both bands energy gap.

  It does not seem possible to provide a general expression for $\epsilon^*$ in
terms of $\Delta_\alpha$  and  an arbitrary inter-band scattering
strength. Still,
in principle, we can evaluate any thermodynamic property of a two-band material
knowing  the solutions $u$ of the system (\ref{b2}) and the gap $\epsilon^*$.

%\section{Penetration depth and its anisotropy}

If the ground state functions (which we call
now $f^{(0)}$, $g^{(0)}$) are known, one can study  perturbations of the
uniform state such as penetration of a weak magnetic field, i.e., the
problem of
the London penetration depth. The perturbations,
$f^{(1)},\,\,g^{(1)}$, should be
found from the full Eilenberger equations; \cite{Eil}  we have for
the first band:
\begin{eqnarray}
{\bm v}{\bm \Pi}f_1 =2\Delta_1 g_1 - 2\omega f_1 +
\frac{ 1}{ \tau _{11}}  [g_1\langle f\rangle_1 -f_1\langle g
\rangle_1]
  \nonumber\\
  + \frac{1}{ \tau _{12}}[g_1 \langle f
  \rangle_2 -f_1\langle g \rangle_2] \,,
\label{eq39}
\end{eqnarray}
Here,  ${\bm v}$ is the Fermi velocity, ${\bm \Pi} =\nabla +2\pi i{\bm
A}/\phi_0$. The second equation is obtained by $1\leftrightarrow 2$.
Two equations for the ``anomalous" functions $f^+$ are obtained from these by
complex conjugation and by ${\bm v}\to -{\bm v}$. \cite{Eil}  The
normalizations
$g_\alpha^2+f_\alpha f_\alpha^+=1$   complete the system.

  We look for solutions in the form
\begin{eqnarray}
f_{\alpha}&=&(f_{\alpha}^{(0)}+f_{\alpha}^{(1)})\,e^{i\theta({\bm r})},
\quad  f_{\alpha}^+=(f_{\alpha}^{(0)}+f_{\alpha}^{(1)+})e^{-i\theta({\bm
r})}\,,\nonumber\\
\label{London_approx}
g_{\alpha}&=&g_{\alpha}^{(0) }+g_{\alpha}^{(1) } \,,\qquad \alpha=1,2
\,.
\end{eqnarray}
where  $f_{\alpha}^{(0)}$ and $g_{\alpha}^{(0)}$ can be expressed in
terms of $u$'s
obtained solving the system (\ref{b2}). The form (\ref{London_approx})
takes into account that in the London approximation only the overall
phase depends
on coordinates. We obtain  for the corrections after some algebra:
\begin{eqnarray}
&&g_1^{(1)}\Delta_1^{\prime}-f_1^{(1)}\omega^{\prime}_1=if_1^{(0)}{\bm v}{\bm
P}/2\,,\nonumber\\
&&g_1^{(1)}\Delta_1^{\prime}-f_1^{(1)+}\omega^{\prime}_1=if_1^{(0)}{\bm v}{\bm
P}/2\,,\label{**}\\
&&2g_1^{(0)}g_1^{(1)}+f_1^{(0)}(f_1^{(1)}+f_1^{(1)+})=0\,,
\nonumber
\end{eqnarray}\\
where ${\bm P}=\nabla\theta+2\pi{\bm A}/\phi_0$ and
\begin{eqnarray}
\Delta_1^{\prime}&=&\Delta_1
+ f_1^{(0)}/2\tau_{11} + f_2^{(0)}/2\tau_{12}\,,
\label{D'}\\
\omega^{\prime}_1&=&\omega
+g_1^{(0)}/2\tau_{11}+g_2^{(0)}/2\tau_{12} \,.\label{w'}
\end{eqnarray}
The equations for the second band (decoupled from the first) are obtained by
$1\leftrightarrow 2$.

To evaluate the penetration depth we turn to the Eilenberger
expression for the current density, \cite{Eil}
\begin{equation}
{\bm j}=-4\pi |e|N(0)T\,\, {\rm Im}\sum_{\omega >0}\langle {\bm v}g\rangle\,,
\label{current}
\end{equation}
and compare it with the London relation
\begin{equation}
\frac{4\pi}{c}  j_i=-(\lambda^2)_{ik}^{-1}
\left(\frac{\phi_0}{2\pi}\nabla\theta+{\bm A}\right)_k\,.
\label{London}
\end{equation}
Here, $(\lambda^2)_{ik}^{-1}$ is the tensor of the inverse squared
penetration depth (proportional to the superfluid density tensor);
summation over
$k$ is implied. We now find
$g_1^{(1)}$ from the system (\ref{**}):
\begin{equation}
g_1^{(1)}=\frac{if_1^{(0)2}\,{\bm v}{\bm P}}{2(\Delta_1^{\prime}f_1^{(0)
}+\omega^{\prime}_1g_1^{(0)})} = i\,\frac{f_1^{(0)2}g_1^{(0)}}
{2 \omega^{\prime}_1}\,{\bm v}{\bm P}\,;
\end{equation}
   $g_2^{(1)}$ is obtained by replacement $1\leftrightarrow 2$.
Substituting $g_{\alpha}^{(1)}$ in Eq.\ (\ref{current}) and
comparing with Eq. (\ref{London}) we obtain:
\begin{eqnarray}
(\lambda^2)_{ik}^{-1}=\frac{16\pi^2e^2N(0)T}{c^2} \sum_{\alpha,\omega}
\nu_\alpha\langle
v_iv_k\rangle_\alpha\frac{f_\alpha^2g_\alpha}{\omega^\prime_\alpha} \,.
\label{lambda}
\end{eqnarray}
Only the unperturbed functions $f,g$ enter the penetration depth; for
brevity we dropped the superscript $(0)$. Equation (\ref{lambda}) is our main
result.   Thus, to evaluate the penetration depth for given order parameters
$\Delta_\alpha$ in the presence of scattering, one has to solve the system
(\ref{b2}) for $u_\alpha(\omega)$, then to substitute the equilibrium functions
$f_\alpha,\,\,g_\alpha$ (given in (\ref{fg})) in Eq.\ (\ref{lambda}) to sum up
over $\omega$.\\

%\subsubsection{Clean limit: all $\tau_{\alpha\beta}\to\infty$}

The band calculations \cite{Bel} yield for MgB$_2$ the following  averages:
$\langle v_a^2\rangle_1=33.2$, $\langle v_c^2\rangle_1=42.2$,
$\langle v_a^2\rangle_2=23$,
and $\langle v_c^2\rangle_2=0.5\times 10^{14}\,$cm$^2$/s$^2$. Tensors
$\langle v_iv_k\rangle_1$ and
$\langle v_iv_k\rangle_2$ have opposite anisotropies:
\begin{equation}
\frac{\langle v_a^2\rangle_1}{\langle v_c^2\rangle_1}\approx 0.79\,,\quad
\frac{\langle v_a^2\rangle_2}{\langle v_c^2\rangle_2}\approx 46.\,,
\end{equation}
whereas   averaging over the whole Fermi surface yields a nearly
isotropic result:
$\langle v_a^2\rangle /\langle v_c^2\rangle \approx 1.2\,$.

\begin{figure}[t]
\includegraphics[angle=0,width=90mm]{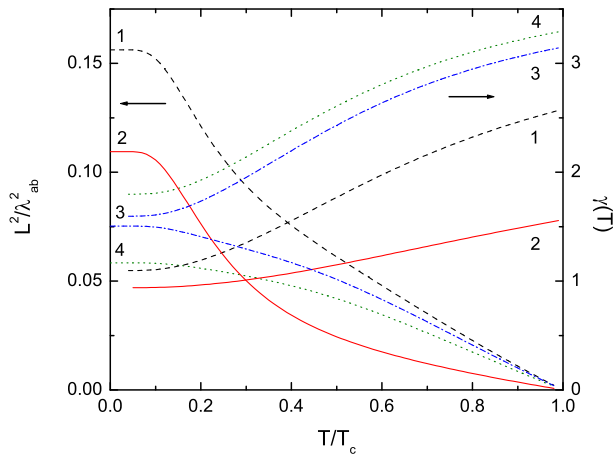}
\caption{\label{f1} The anisotropy $\gamma_\lambda=\lambda^2_{c}\lambda^2_{ab}$
and the inverse square of the penetration depth
$L^2/\lambda^2_{ab}$ {\it versus} $T/T_c$;   $L^2=16\pi^2e^2N(0)\langle
v_a^2\rangle/c^2$.
  The curves labelled 1 is the clean limit, all
$1/\tau$ are zero. The curves labelled 2 and 3 are calculated for a
weak inter-band
scattering: $\tau_{12}\Delta_1(0)=500$, $\tau_{21}\Delta_2(0)=2000$
($\hbar=1$);
the curve 2 is for   a clean $\pi$-band, $\tau_{11}\Delta_1(0)=10$, and a
dirty $\sigma$, $\tau_{22}\Delta_2(0)=0.1$; the curve 3 is for a
dirty $\pi$ and
clean $\sigma$: $\tau_{11}\Delta_1(0)=0.1$, $\tau_{22}\Delta_2(0)=10$. Curves 4
are for the intermediate inter-band scattering strength
$\tau_{12}\Delta_1(0)=5$,
$\tau_{21}\Delta_2(0)=20$, and $\tau_{11}\Delta_1(0)=0.05$,
$\tau_{21}\Delta_2(0)=2 $.}
\end{figure}

In the {\it clean} limit (all $\tau_{\alpha\beta}\to\infty$)
$\omega^\prime=\omega$ and
$\Delta^{\prime}_\alpha=\Delta_\alpha$. Besides,
$u_\alpha=\omega/\Delta_\alpha$
and $f_\alpha^2g_\alpha/\omega^\prime_\alpha=
\Delta_\alpha^2/(\omega^2+\Delta_\alpha^2)^{3/2}$. Expression (\ref{lambda})
reduces to the  result given in Ref. \onlinecite{Kogan}.
For MgB$_2$, it gives nearly isotropic penetration depth at low temperatures:
at $T=0$  the sums over $\omega$ in Eq.\ (\ref{lambda}) are the same;
this gives
   $\gamma_\lambda(0)=\lambda_{cc}/\lambda_{aa}=\sqrt{\langle
v_a^2\rangle /\langle v_c^2\rangle}\approx 1.1$. Near $T_c$, the sums  are
$\propto\Delta_\alpha^2$, and the contribution of the  strongly anisotropic
$\sigma$-band with the large gap  dominates; this gives $\gamma_\lambda(T_c)
  \approx 2.6\,$. The  curve 1 in
Fig.\ \ref{f1} shows $\gamma_\lambda(T)$ for this case. \\

%\subsubsection{Zero inter-band scattering ($\tau_{12}=\tau_{21}=\infty$)}

{\it Zero inter-band scattering ($\tau_{12}=\tau_{21}=\infty$)}. If only the
intra-band scattering is present,  the functions $f,g$ are the same as in the
clean limit. We readily obtain:
\begin{equation}
\frac{f_\alpha^2g_\alpha}{\omega^\prime_\alpha}= \frac{ \Delta_\alpha^2
}{\beta_\alpha^2(\beta_\alpha + 1 /2\tau_{\alpha\alpha}) }\,,
\label{1/t12=0}
\end{equation}
with $\beta_\alpha^2=\omega^2+\Delta_\alpha^2$. This expression
appears in the
standard  penetration depth calculations, see e.g. Ref. \onlinecite{AGD}. For
known $\Delta_\alpha(T)$, the sums in Eq. (\ref{lambda}) can be evaluated
numerically; however, for
$T\to 0,\,\,T_c\,$,  and in the dirty limit they can be done analytically.

At $T=0$, the sums are replaced with integrals according to $2\pi
T\sum_\omega=\int_0^\infty d\omega$. Denoting ${\cal I}(T )=2\pi T\sum_\omega
\Delta ^2/ \beta ^2(\beta  + 1 /2\tau )$ we obtain ${\cal I}(0)=1$  for
$\tau\to\infty$ and ${\cal I}(0 )= \pi\tau\Delta$ for $\tau\Delta\ll 1$.

Near $T_c$, $g\,$$\to$$ 1$ and we have for clean bands $\sum_\omega
f^2/\omega^\prime =7\zeta(3)\Delta^2/8\pi^3T_c^3$, whereas for dirty
bands it is
$\tau\Delta^2/4T_c^2 $.

Different impurities introduced to MgB$_2$  may affect differently
the scattering
within the bands. \cite{Maz1,selective_scattering}  It is
  of interest to see how the anisotropy of $\lambda$ is affected by
differences in scattering times $\tau_{11}$ and $\tau_{22}$. We first
look at two
limiting situations when one of the bands is clean while the other is dirty. If
the first ($\pi$) band  is clean and the second ($\sigma$) is a dirty extreme
($\tau_{22}\Delta_2\to 0$), one can disregard the contribution of the
dirty band to
obtain for both $T=0$ and
$T=T_c$:
\begin{equation}
\gamma_\lambda (0) =\gamma_\lambda (T_c) \approx  \sqrt{\frac{ \langle
v_a^2\rangle_1} {
\langle v_c^2\rangle_1} }\approx 0.89  \,.
\label{gamma_clean-dirty-1}
\end{equation}
If the ($\pi$) band  is dirty and the ($\sigma$) is clean, we have
\begin{equation}
\gamma_\lambda (0) =\gamma_\lambda (T_c) \approx  \sqrt{\frac{ \langle
v_a^2\rangle_2} {
\langle v_c^2\rangle_2} }\approx 6.8  \,.
\label{gamma_dirty_clean}
\end{equation}
These two numbers constitute the minimum and maximum possible values for
$ \lambda$-anisotropy of MgB$_2$. Thus, when one of the bands is clean and the
other is dirty we expect a weakly $T$ dependent $\gamma_\lambda$, the value of
which is determined by the clean band.

  If the intra-band scattering is strong in both bands
($\tau_{11}\Delta_1\sim\tau_{22}\Delta_2\ll 1$, $\tau_{12}=\infty$), the bands
contribute to the superfluid density tensor $(\lambda^2)_{ik}^{-1}$ as two
independent dirty superconductors. To see   this, we note that
$\omega_1^\prime\approx g_1/2\tau_{11}$ and the sums over $\omega$ in
Eq, (\ref{lambda}) can be evaluated exactly:
\begin{equation}
\sum_\omega\frac{f_1^2g_1}{\omega^\prime_1}
=\sum_\omega\frac{2\tau_{11}\Delta_1^2}{\omega^2+\Delta_1^2}=\frac{
\tau_{11}\Delta_1}{2 T}\tanh\frac{\Delta_1}{2T}\,.
\end{equation}
Then, we arrive at the result obtained by Gurevich with the help of
the dirty limit
Usadel equations: \cite{Sasha}
\begin{equation}
(\lambda^2)_{ik}^{-1}=\frac{4\pi^2 }{c^2\hbar}
\sum_\alpha\sigma_{ik}^{(\alpha)}\Delta_\alpha\tanh\frac{\Delta_\alpha}{2T}
\label{lambda-dirty}
\end{equation}
where the anisotropic conductivities of the two bands $\sigma_{ik}^{(\alpha)}=2
e^2\langle v_iv_k\rangle_\alpha \tau_{\alpha\alpha}\nu_\alpha N(0)$
are introduced
(we write here $\hbar$ explicitly to avoid confusion in dimensions).
This yields:
\begin{equation}
\gamma_\lambda^2 (0)=  \frac{\sigma_{aa}^{(1)}\Delta_1
+\sigma_{aa}^{(2)}\Delta_2  }{
\sigma_{cc}^{(1)}\Delta_1 +\sigma_{cc}^{(2)}\Delta_2}   \,\,,
\label{gamma_dirty-1}
\end{equation}
\begin{equation}
\gamma_\lambda^2 (T_c)=  \frac{\sigma_{aa}^{(1)}\Delta_1^2
+\sigma_{aa}^{(2)}\Delta_2^2  }{
\sigma_{cc}^{(1)}\Delta_1^2 +\sigma_{cc}^{(2)}\Delta_2^2}   \,\,.
\label{gamma_dirty-1-Tc}
\end{equation}\\

%\subsubsection{Strong inter-band scattering}

Finally, we discuss the possibility of a  {\it strong inter-band
scattering}. As was shown above, in this case
$u=\omega/\epsilon^*+{\cal O}(1/\zeta)$ in both bands, see Eqs. (\ref{strong
inter}) and (\ref{v1,strong}). Then, the Eilenberger functions are
also the same
in the two bands:
$f= \epsilon^*/\sqrt{\omega^2+\epsilon^{*2}}$,  $g=
\omega/\sqrt{\omega^2+\epsilon^{*2}}$.
Evaluation of the sums over $\omega$ in Eq. (\ref{lambda}) is now simple:
\begin{eqnarray}
(\lambda^2)_{ik}^{-1}=\frac{8\pi^2e^2N(0) }{c^2}\,
\epsilon^*\tanh\frac{\epsilon^*}{2T}\sum_\alpha
\nu_\alpha\langle v_iv_k\rangle_\alpha\tau_\alpha
\label{lambda_dirty}
\end{eqnarray}
where $\epsilon^*(T)$ is given in Eq. (\ref{GL_limit}) and
\begin{equation}
\tau_1=\frac{\tau_{11}\tau_{12}}{\tau_{11}+\tau_{12}}\,,\quad
\tau_2=\frac{\tau_{22}\tau_{21}}{\tau_{22}+\tau_{21}}\,.
\end{equation}
Thus, all components of $(\lambda^2)_{ik}$ have the same $T$
dependence and the anisotropy parameter is $T$ independent:
\begin{equation}
\gamma_\lambda^2 =\frac{\nu_1\langle v_a^2\rangle_1\tau_1
+\nu_2\langle v_a^2\rangle_2\tau_2 }{\nu_1\langle v_c^2\rangle_1\tau_1
+\nu_2\langle v_c^2\rangle_2\tau_2}\,.
\label{gam_dirty}
\end{equation}

If all $\tau$'s are the same, we have:
\begin{equation}
  \gamma_\lambda^2 = \langle v_a^2\rangle
  \big/\langle v_c^2\rangle  \,.
\label{gam_dirty_dirty}
\end{equation}
For $T\to T_c$, this result  was obtained in Ref. \onlinecite{Kogan}; we
now see that it holds at any temperature.

To recover the behavior of  $\gamma_\lambda(T)$   between $0$
and $T_c$ one needs explicit dependencies $\Delta(T)$. Qualitatively, this
behavior can be studied assuming
$\Delta_\alpha(T)\approx\Delta_\alpha(0)(1-t^2)$
with, e.g., $\Delta_2(0)=4\Delta_1(0)=2 T_c$.  Figure 1 shows results of
numerical evaluation of $\gamma_\lambda(T)$ for scattering parameters
given in the
caption (which are not that extreme as in the above discussion).
The curves $\gamma_\lambda(T)$ are obtained by solving Eqs.
(\ref{b2}) for $u$'s in
two bands and then by evaluation of the sums in Eq. (\ref{lambda}). It is worth
noting that although the $T$ dependences shown in the figure are obtained using
approximate $\Delta(T)$, the end points of these curves at $T=0$ and
$T=T_c$ are
exact.

  We conclude  that both the inter- and intra-band scattering affect
strongly the superconducting anisotropy of two-band superconductors in
general and of MgB$_2$, in particular. If one of the MgB$_2$ bands is
dirty, the
anisotropy is dominated by a cleaner band:
$\gamma_\lambda(T)$ is close to unity (and might be even less than 1)
if the $\pi$
band is in the clean limit, whereas in the opposite situation of a
clean $\sigma$,
$\gamma_\lambda(T)$ is large being in both cases weakly $T$ dependent. The
inter-band scattering  suppress the $T$ dependence of $\gamma_\lambda$ as
compared to the clean limit discussed earlier. \cite{Kogan}

We thank J. Clem and S. Bud'ko  for useful discussions.
   Ames Laboratory is operated for the U. S. Department of Energy by
Iowa State University under Contract No. W-7405-Eng-82.

\end{document}